\documentclass[fleqn, default]{sn-jnl}

\usepackage{hyperref}
\usepackage{latexsym}
\usepackage{epic,eepic}

\usepackage{graphicx}%
\usepackage{multirow}%
\usepackage{amsmath,amssymb,amsfonts}%
\usepackage{amsthm}%
\usepackage{mathrsfs}%
\usepackage[title]{appendix}%
\usepackage{xcolor}%
\usepackage{textcomp}%
\usepackage{manyfoot}%
\usepackage{booktabs}%
\usepackage{algorithm}%
\usepackage{algorithmicx}%
\usepackage{algpseudocode}%
\usepackage{listings}%

\newtheorem{definition}{Definition}
\newtheorem{proposition}{Proposition}
\newtheorem{theorem}{Theorem}

\newcommand{\ConfigType}{\textit{ConfigType}}

\newcommand{\Seq}[1]{\langle #1 \rangle}
\newcommand{\CRule}[4]{(#1,#2,\Seq{#3},#4)} 

\newcommand{\False}{{\mit f\!alse}}

\renewcommand{\And}{\,\wedge\,}
\newcommand{\Or}{\,\vee\,}
\newcommand{\Equiv}{\leftrightarrow}
\newcommand{\Impl}{\rightarrow}
\newcommand{\Sat}{\models}
\newcommand{\Satisfiable}[1]{sat(#1)}
\newcommand{\Param}[1]{{\tt #1}}
\newcommand{\HasType}{\!:\!}
\newcommand{\Lang}[1]{{\cal L}(#1)}
\newcommand{\Logic}{L}
\newcommand{\LogicOver}[1]{L_{#1}}

\newcommand{\Denot}[1]{[\![#1]\!]}

\newcommand{\IffArrow}{\, \Leftrightarrow \,}
\newcommand{\Implies}{\Rightarrow}

\usepackage{array}
\newcolumntype{P}{>{\raggedright}p}
\newcolumntype{L}{>{$}l<{$}}
\newcolumntype{C}{>{$}c<{$}}
\newcolumntype{R}{>{$}r<{$}}
\newenvironment{BNFarray}[1][c]{\ignorespaces
  \begin{array}[#1]{ll>{}L}}{\end{array} \ignorespacesafterend}

\newcommand{\BNFtop}[2]{\multicolumn{2}{l}{
    #1\;\;\BNFdef}&\textit{#2}}

\newcommand{\BNFrow}[3][0pt]{\\[#1]\multicolumn{2}{l}{\hspace{1em}#2}&\hspace{1em}#3}
\newcommand{\BNFsep}{\;\;\pipe\;\;}
\newcommand{\pipe}{\rule[-.4ex]{0.08em}{2ex}}

\newcommand{\Syntax}[1]{#1}
\newcommand{\BNFdef}{\mathord:\mathord:\mathord=}

\newcommand{\KWord}[1]{{\tt #1}}
\newcommand{\BNFnonterm}[1]{\mbox{\textit{#1}}}
\newcommand{\BNFterm}[1]{\mbox{\textbf{#1}}}

\newcommand{\Na}[1]{\mathtt{#1}}
\newcommand{\Co}[1]{\mathtt{\overline{#1}}}

\newcommand{\rest}[1]{\mbox{$\setminus #1$}}

\newcommand{\Def} {\stackrel{{\rm def}}{=}}

\newcommand{\IfCCS}{\mbox{\hspace{1pt}if }}
\newcommand{\Then}{\mbox{ then }}
\newcommand{\ElseCCS}{\mbox{ else }}

\begin{document}

\title{A Logical Way to Negotiate Services}

\author[1]{Glenn Bruns}
\author[2]{Mauricio Cortes}

\affil[1]{\orgdiv{School of Computing and Design}, \orgname{California State University, Monterey Bay}, \orgaddress{\street{100 Campus Center}, \city{Seaside}, \postcode{93955}, \state{California}, \country{USA}}}

\affil[2]{\orgname{Joyent}, \orgaddress{\street{645 Clyde Avenue, Suite 502}, \city{Mountain View}, \postcode{94043}, \state{California}, \country{USA}}}

\abstract{
  Service providers commonly provide only a fixed catalog of
  services to their clients.  Both clients and service providers
  can benefit from service negotiation, in which a client makes
  a query for a specific service, and the provider counters with
  an offer.  The query could include parameters that control the
  performance, reliability, and function of the service.
  However, a problem with service negotiation is that it can
  be expensive for a service provider to support.

  In this paper we define a formal negotiation policy language that
  enables automated service negotiation.
  In the model supported by the language, service providers
  can recursive obtain the services they need from sub-providers.
  The queries made by clients, and the offers
  returned from service providers, are expressed in quantifier-free
  first-order logic.  Quantifier elimination is used to transform
  constraints between providers and sub-providers.
  The pattern of interaction between clients
  and service providers is defined in process algebra.
  We show a correctness
  property of our language: if sub-providers respond
  positively to queries, then so does the provider itself.
}

\maketitle

\section{Introduction}

Service providers -- such as internet service providers, wireless
service providers, storage service providers, and providers of
specialized online services -- typically provide a static
catalog of services.  As a simple example, an internet service
provider might offer two options: 100 Mbps down and 10 Mbps up,
or 50 Mbps down and 2 Mbps up.  While the simplicity of this
``service catalog" approach is helpful, it lacks flexibility. 
For example, one customer might need a download speed of at least 300 Mbps
but an upload speed of only 10 Mbps, while another customer might be happy 
with an download speed of 50 Mbps and an upload speed of 20 Mbps.  The lack of flexibility leads to lost opportunities for service providers and low
perceived value for customers.

In a more flexible approach, a client could negotiate with a
service provider to obtain the service that fits her needs.
However, this kind of flexibility is commonly only available
for important clients and involves expensive manual work by 
the service provider.

A solution to the problem would be a framework for service negation
that provides flexibility but reduces costs through automation.  Here we define
such a framework.  The main features of our framework are as
follows:

\begin{enumerate}
    \item Services have a hierarchical structure.  A service provider
    may depend on ``sub-providers".  For example, the provider of
    a service to edit and compile LaTeX documents might pay sub-providers
    for storage, computation, and payment services.  Thus, a service
    provider can also play the part of client with respect to other
    service providers.

    \item A service provider receives a {\em query} from a client that
    defines constraints on the service needed by the client.
    For example, a query to an internet service provider might specify
    a minimum download speed of 100 Mbps, a minimum upload speed of
    20 Mbps, and a maximum price of 75 USD per month.
    The service provide responds to the client with an {\em offer} that
    defines the service that can be provided.  Both queries and offers
    are defined as quantifier-free formulas of first-order logic.

    \item If a service provider depends on sub-providers, then responding
    to a query will require sending sub-queries to the sub-providers, and
    then combining the received sub-offers to create a top-level offer.
    Also, a sub-offer from one sub-provider can affect sub-queries sent to other
    sub-providers.
    For example, a document processing provider may need a certain
    amount of storage, which can be provided by two sub-providers.
    If the first sub-provider can provide most of the needed storage,
    the query to the second provider can request less storage.

    \item The negotiation policy of a service provider is captured in
    a formal policy that defines the service parameters, constraints
    on the services that can be offered, and relationship between
    services provided by sub-providers and the top-level service that
    is provided.

    \item If service providers define their negotiation policies, then
    negotiation can be automated.  The pattern of interaction between
    clients and providers is defined in process algebra.  
\end{enumerate}

In a simple running example used throughout this paper, a storage
provider is a broker that obtains storage from two other
storage providers.  The negotiation policy of the top-level
provider might specify that the parameters of the service are
the amount of storage (in GBytes) and the yearly cost of service
(in USD), that the storage offered is the sum of the storage offered
by the sub-providers, and that the price offered includes a 10\%
markup over the cost of the storage obtained from the sub-providers.

Suppose the storage provider receives a query for 10 GBytes of
storage at a price of 5 USD/year.  The storage provider then sends
a query to sub-provider 1 for 10 GBytes of storage at a price of
4.55 USD (a 10\% markup on 4.55 USD gives 5.00 USD).  Suppose an 
offer is received for 5 GBytes of storage
at a price of 4.20 USD.  The storage provider would then send
a query to sub-provider 2 for 5 GBytes of storage at a price of
4.90 USD.  At this price the top-level provider can still obtain
a 10\% markup on the combined storage from the two sub-providers.
The top-level provider than makes an offer of 10 GBytes of storage
at a price of 5 USD/year.

In this example, negotiation of the storage provider with the
two sub-providers takes place sequentially.  Later in the paper
we describe both sequential and parallel patterns of interaction
with sub-providers.

The main contribution of our work is the formal definition of a
language for service negotiation.  We define the syntax and semantics
of the language, and show an important formal property of negotiation
behavior: if sub-negotiators respond positively to queries, then so
does the top-level negotiator.
We also define several extensions to the policy language, including
support for making parallel queries to sub-providers.

In the following section of the paper we briefly review our
hierarchical service negotiation model, which was presented
in \cite{BC2011a}.  In Sections \ref{sec:policy} and \ref{sec:interpretation},
we define the syntax and semantics of the the policy language.
In Section \ref{sec:properties}, we define and
prove a correctness condition of the language.  In
Section \ref{sec:extensions} several extensions to the
language are discussed, and an implementation of the
language is described in Section \ref{sec:implementation}.
The last two sections of the paper describe related work,
and offer concluding remarks.

\section{Hierarchical Service Negotiation}
\label{sec:hierarchical}

In hierarchical service negotiation, the negotiation process has this form:

\begin{enumerate}
  \item The client makes a request to a negotiation server to
        initiate negotiation.
  \item The negotiation server acknowledges, returning the
        terms of negotiation, which identifies service parameters
        and the constraints on them.
  \item The client sends the negotiation server a query, which is
        a condition over the service parameters.
  \item The negotiation server makes queries, and obtains
        offers, from one or more ``sub''-negotiation servers.
  \item The negotiation server sends an offer to the client.
  \item This query/offer process is repeated until either the
        client accepts an offer, in which case the negotiation
        server returns an invoice, or one of the two parties
        terminates negotiation.
\end{enumerate}
The details of this process are explained in \cite{BC2011a},
which includes a description of a negotiation protocol.  Not
every negotiation server will contact sub-servers -- some
servers must be ``base cases'' that make offers without
contacting sub-servers.  However, our focus here is in the
``inductive case''.

In this section we define how the terms of negotiation are
defined, and the kinds of formulas used for queries and
offers.  Please note that in what follows negotiation
servers are sometimes referred to as ``negotiators'',
and sometimes simply as ``servers''.

\subsection{Configuration Types}
\label{subsec:config-types}

A {\em configuration type} (``config. type'' for short)
defines the negotiable parameters of a service,
their types, and any constraints over the parameters.  The
syntax of config. types is defined in
Fig.~\ref{fig:config-types}.  For example, the config. type
of a simple storage service might be
$\{\Param{capacity} \HasType decimal,\,
   \Param{price} \HasType decimal;\, 
\Param{capacity} \geq 0 \And \Param{price} \geq 0\}$.  
Here every parameter has basic type ``decimal''; later in
the paper we discuss support for additional types.
We write $\Lang{ct}$ for the set of parameter names
appearing in configuration type $ct$, and $\ConfigType$ for
the set of all configuration types.

\begin{figure}
\[
\begin{array}{lcl}
   \BNFnonterm{ct} & \BNFdef & \{\BNFnonterm{params}; \phi\}\\
   \BNFnonterm{params} & \BNFdef & \epsilon \BNFsep \BNFterm{id} \HasType \BNFnonterm{bt}, \BNFnonterm{params}\\
   \BNFnonterm{bt} & \BNFdef & \KWord{decimal}
\end{array}
\]
\caption{The abstract syntax of configuration types, where $id$ ranges over
  identifiers, and $\phi$ is a formula such that all free variables appearing in $\phi$ are in $\Lang{ct}$.}
\label{fig:config-types}
\end{figure}

\subsection{Linear Constraints}

To express queries and offers on a service, as well
as constraints in config. types, we use a
first-order logic in which the atomic predicates are
conditions on parameters of config. types.  The
abstract syntax of the logic we used is defined in
Fig.~\ref{fig:logic-def}.  For example, a formula describing
a condition on a storage service is $\Param{capacity} \geq
10 \And \Param{price} \leq 5$, where capacity is given in
GBytes and price is a yearly price in dollars.

\begin{figure}
\[
\begin{BNFarray}
   \BNFtop{t}          {term}
   \BNFrow{x}                           {variable}
   \BNFrow{c}                           {constant}
   \BNFrow{t_1 \Syntax{+} t_1}          {addition}
   \BNFrow{c \Syntax{\ast} t}           {multiplication by a constant}
\end{BNFarray}
\begin{BNFarray}
   \BNFtop{\phi}       {formula}
   \BNFrow{t_1 \sim t_2}                {atomic predicate}
   \BNFrow{\phi \Syntax{\And} \phi}     {conjunction}
   \BNFrow{\Syntax{\neg} \phi}          {negation}
   \BNFrow{\exists x \Syntax{.} \phi}   {existential quantification}
\end{BNFarray}
\]
\caption{The abstract syntax of terms and formulas, where $x$ ranges
over variable symbols, and $\sim$ ranges over $\{=,\neq,\leq,\geq,<,>\}$.}
\label{fig:logic-def}
\end{figure}

We write $\Logic$ for the set of formulas generated by this
syntax.  As usual, other logical operators, like $\Or$,
$\forall$, and $\Impl$, are derived.  The notion of free
variables of a formula is assumed to be understood.  If all
free variables appearing a formula are elements of
$\Lang{ct}$ for some config. type $ct$, then we say $\phi$
is a formula {\em over} $ct$.  The set of all formulas over
$ct$ is written $\LogicOver{ct}$.  For example, letting $ct$
be the example config. type defined above for a storage
service, a formula in $\LogicOver{ct}$ is $\Param{capacity} = 2 
\And \Param{price} \leq 5$.

A query over a service with config. type $ct$ is then
defined to be a quantifier-free formula of $\LogicOver{ct}$.
An offer is also a quantifier-free formula of
$\LogicOver{ct}$.  When a client accepts an offer, it
provides a quantifier-free formula of $\LogicOver{ct}$ that
logically implies the offer, and that specifies values
for all parameters of $ct$.  For example, in the storage
example, accepting an offer $\Param{capacity} = 2 \And
\Param{price} \geq 5$ might be the formula $\Param{capacity} = 2
\And \Param{price} = 5$.


The logic we use defines linear constraints -- in other
words, systems of linear inequalities.  When this logic is
used over the domain of the naturals, it is referred to as
{\em Presburger Arithmetic}
\cite{Presburger1930,Presburger1991}.  The key property we
require of the logic is that it is possible to compute, from
an arbitrary formula, an equivalent, quantifier-free
formula.  Algorithms for quantifier elimination exist for
this logic when it is interpreted over the naturals, the
integers, the reals, or the rationals.  For example,
Fourier-Motzkin elimination \cite{schrijver1998theory} can
be used when the logic is interpreted over the reals or the
rationals.  In what follows we write $QF(\phi)$ for a
formula that is logically equivalent to PA formula $\phi$, but
quantifier-free.

Intuitively, the use of existential quantification can be
viewed as a way to project a formula with free variables
onto some of the variables it contains.  For example, let
$\phi$ be $x < 5 \And x > y \And y > 0$.  To see what $\phi$
says about $x$, we ``quantify away'' the $y$ in $\phi$, to
get $\exists y.(x < 5 \And x > y \And y > 0$.  Applying
quantifier elimination, we get $QE(\phi) = x < 5 \And x >
0$.  Intuitively, this formula expresses how $\phi$
constrains $x$.

We interpret logical formulas in the usual way.  Briefly, an
{\em interpretation} $I$ consists of a non-empty domain of
objects, plus interpretations of constant, function, and
predicate symbols.  The logic we use has
$\{=,\neq,\leq,\geq,<,>\}$ as its predicate symbols, and $+$
and $\ast$ as its function symbols.  A {\em valuation} $v$
for an interpretation maps variable symbols to elements of
the interpretation's domain.  In this paper we interpret
formulas over the rationals, and adopt the usual arithmetic
interpretation of the function and predicate symbols.

We write $I,v \Sat \phi$ if a formula $\phi$ of $\Logic$, possibly
containing free variables, is satisfied by interpretation
$I$ under valuation $v$.  We write $v \Sat \phi$ if $I,v
\Sat \phi$ where $I$ is the expected interpretation that was
just described.  We write $\Sat \phi$ if $\phi$ is {\em
  logically valid}; i.e. is satisfied by all $I$ and $v$.
Also, we also write $\Satisfiable{\phi}$ to mean that $\phi$
is satisfied by some $I$ and $v$ -- in other words, $\not
\Sat \neg \phi$.  We write $\phi \Implies \psi$ if whenever
$I,v \Sat \phi$ then $I,v \Sat \psi$.  We write $\phi
\IffArrow \psi$ if $\phi \Implies \psi$ and $\psi \Implies
\phi$.

\section{A Negotiation Policy Language}
\label{sec:policy}

In this section we define the syntax of our negotiation
policy language.  Informally, a policy consists of a name, a
list of the negotiators it refers to, the config. type it
supports, and a collection of rules.  Each rule defines a
relationship between the negotiation service being supported
and the negotiation services being used.

Fig.~\ref{fig:storage-example-1} shows a simple 
negotiation policy for a storage service provider.  The
provider is acting as a broker, obtaining storage from
two other providers.  Informally, the policy says that
sub-providers $s1$ and $s2$ are used, and that the offer
made by the provider is related to the offers from the
sub-providers in the following way: the capacity offered
is the sum of the capacities offered by $s1$ and $s2$,
and the price offered is 10\% more than the sum of the
prices offered by $s1$ and $s2$.  

The ``serves'' part specifies the config. type supported 
by the policy.  In this example, 'storage' is defined to
be the configuration type shown in Section \ref{subsec:config-types}:
$\{\Param{capacity} \HasType decimal,\,
   \Param{price} \HasType decimal;\, 
\Param{capacity} \geq 0 \And \Param{price} \geq 0\}$.  

The single rule of the policy specifies how the service is
provided.  The ``trigger'' part of the rule is a condition
on queries from clients.  The rule is ``applicable" if the
conjunction of the trigger condition and the query are
satisfiable.  In this example no condition is imposed on
queries.

The ``uses'' part of the rule shows that
storage negotiation servers with names $s1$ and $s2$ are to
be used as sub-negotiators.  

The ``offer'' part of the rule
specifies how parameters of the offered service depend on
parameters of the sub-services.  In this example, the
offered capacity is the sum of the capacities of the
sub-services, and the offered price is the sum of the prices
of the sub-services, plus a $10\%$ markup.  

The ``constraint'' part of the rule is a condition that must
hold between the service provided by the policy and the
negotiated sub-services.  In this example no constraint is
used, but the constraint could be used to specify, for
example, that each sub-service provide half of the total
storage.

\begin{figure}
{\small
\begin{verbatim}
policy storage-brokering {
  serves: storage
  rules: [
   { trigger: true,
     uses: [s1:storage, s2:storage],
     offer: { capacity := s1.capacity + s2.capacity,
              price := 1.1*(s1.price + s2.price) },
     constraint:true},
  ]
}
\end{verbatim}}
\caption{A Negotiation Policy for a Storage Brokering Service.}
\label{fig:storage-example-1}
\end{figure}

Fig.~\ref{fig:policy-syntax} defines the syntax of
policies.  We use an extended version of BNF in which $List(\beta)$
indicates zero or more repetitions of the phrases defined by
grammar symbol $\beta$ enclosed in square brackets and
separated by commas.  Also, $Struct(\beta_1,\ldots,\beta_n)$
indicates the phrases defined by symbols
$\beta_1,\ldots,\beta_n$ enclosed within curly brackets.

\begin{figure}
\[
\begin{array}{lcl}
  \BNFnonterm{policy}          & \BNFdef & \KWord{policy} \; id \; 
                                 Struct(\BNFnonterm{serves-part}, \BNFnonterm{rules})\\
  \BNFnonterm{serves-part}     & \BNFdef & \KWord{serves:}\; \BNFnonterm{type-expr}\\
  \BNFnonterm{rules}           & \BNFdef & \KWord{rules:}\; List(\BNFnonterm{rule})\\
  \BNFnonterm{rule}            & \BNFdef & 
                                 Struct(\BNFnonterm{trigger-part}, \BNFnonterm{uses-part},
                                 \BNFnonterm{offer-part}, \BNFnonterm{constraint-part})\\
  \BNFnonterm{trigger-part}    & \BNFdef & \KWord{trigger:}\; \phi\\
  \BNFnonterm{uses-part}       & \BNFdef & \KWord{uses:}\; List(id : \BNFnonterm{type-expr})\\
  \BNFnonterm{offer-part}      & \BNFdef & \KWord{offer:}\; \BNFnonterm{assn}\\
  \BNFnonterm{constraint-part} & \BNFdef & \KWord{constraint:}\; \phi\\
  \BNFnonterm{type-expr}       & \BNFdef & id \BNFsep \BNFnonterm{ct}\\
  \BNFnonterm{assn}            & \BNFdef & List(id \; \BNFterm{:=} \; t)
\end{array}
\]
\caption{The syntax of policies, where $id$ is an identifier, $ct$ is a configuration type, $t$ is a term of $L$, and $\phi$ is a formula of $L$.}
\label{fig:policy-syntax}
\end{figure}

If a policy $p$ serves config. type $ct$, then the trigger
of every rule must be a formula over $ct$.  The constraint of
every rule must be a formula over $ct,ct_1,\ldots,ct_n$, where
$ct_i$ is the config. type of negotiation server $s_i$
appearing in the ``uses" part of the rule.  In an {\em assignment}
(nonterminal $\BNFnonterm{assn})$, every $id$ must be a parameter
of $ct$, and the variables appearing in every term $t$ must be
parameters of config. types $ct_1,\ldots,ct_n$.

In the policy example of Fig.~\ref{fig:storage-example-1},
server prefixes are used to distinguish parameters
associated with different negotiation servers.  This feature
is supported in our language, and is used in examples, but
for simplicity we will not support this feature in formal
definitions.

\section{Interpreting Negotiation Policy}
\label{sec:interpretation}

In this section we define the meaning of policies.  The
basic idea is that a policy is interpreted as a process that
has a port for accepting queries, a port for returning
offers, and ports for interacting with sub-negotiators.

\begin{figure}
\begin{center}
\includegraphics[scale=.3]{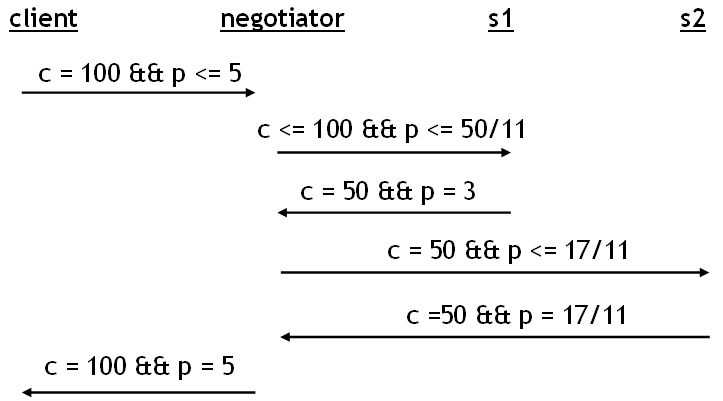}
\end{center}
\caption{Negotiation based on the policy of Fig.~\ref{fig:storage-example-1}}
\label{fig:example1-negot}
\end{figure}

To get intuition for what follows, look at
Fig.~\ref{fig:example1-negot}, a message sequence diagram
showing how the process derived from the policy of
Fig.~\ref{fig:storage-example-1} might behave.  The
negotiator process accepts query $c = 100 \And p \leq 5$ on
its input port (in the figure $c$ is used for
$\Param{capacity}$ and $p$ for $\Param{price}$.)  From this
query, the query to $s_1$ is formed in two steps.  First the
``offers'' part of the policy is used to relate the query to
a condition on the storage needed from $s_1$ and $s_2$.
Then this condition is ``projected onto'' the parameters of
$s_1$ itself by quantifying existentially, and then
eliminating the quantifier.  The query to $s_1$ contains $c
\leq 100$ because sub-negotiator $s_2$ can provide any
memory not provided by $s_1$.

The offer returned from $s_1$ is $c = 50 \And p = 3$.  This
offer must be used in computing the query for $s_2$.  The
offer returned from $s_2$ is $c = 50 \And p \leq 17/11$.
The offer sent to the client is formed by combining the two
sub-offers and the ``offers'' part of the policy, and then
projecting the result onto the parameters of the storage
service, again using existential quantification.  In this
example the offer sent to the client exactly determines
values for all parameters of the service being negotiated,
but in general an offer is any formula over the
configuration type of the service being negotiated.

We define a {\em negotiator} as a process with an interface 
consisting of port $\Na{in}$, which accepts a query, and a
port $\Na{out}$, which produces an offer.  In what follows
we use the process algebra CCS \cite{Milner89}
to describe processes.

\subsection{Core policy language}

Rather than defining the meaning of policies directly, we
use the ``core language approach''.  In this approach a
minimal language is used for defining the language
semantics, and the full language is defined by translation
to the core language.

The syntax of the core language is as follows.  A policy is
either a (core language) rule, or the composition $p_1
\oplus p_2$ of two policies, which are themselves expressed
in the core language.  A rule has the form
$\CRule{ct}{\phi}{s_1:ct_1,\ldots,s_n:ct_n}{\psi}$, where
$ct$ is the config. type of the rule, $\phi$ is the
trigger condition of the rule, $s_1:ct_n,\ldots,s_n:ct_n$
are the names and types of the negotiation servers used in
the rule, and $\psi$ is the condition of the rule.

The translation from the full language to the core language
is straightforward, and so we only sketch it here.  Let $p$
be a policy in the full language, containing rules $r_1,\ldots,r_n$.
From $p$ we derive $r'_1 \oplus \cdots \oplus r'_n$, where
$r'_i$ is the core language form of $r_i$.  The config. type of
every $r'_i$ is the config. type of $p$.  The trigger
condition of $r'_i$ is the trigger of $r_i$.  The servers of
$r'_i$ are the servers of $r_i$.  Finally, the condition $\psi$
of $r'_i$ is the conjunction of the constraint of $r_i$ and
the formula derived from the assignment of $r_i$.  A formula
is obtained from an assignment simply by replacing each
assignment symbol $:=$ with logical symbol $\Equiv$.

\subsection{Semantics of a rule}
\label{sec:rule-semantics}

We now define the semantics of policies, which are
interpreted as processes.  We begin with the case of a
policy that is a single rule.  Informally, the process
denoted by a rule will accept query $q$, query each negotiation
server mentioned in the rule, awaiting a response before querying
the next server, and finally output an offer.  The query for
each server incorporates responses from previous servers.

Let $r$ be a rule
$\CRule{ct}{\phi}{s_1:ct_1,\ldots,s_n:ct_n}{\psi}$ in the core
language syntax.  Then the process $\Denot{r}$ denoted by
$r$ is defined as follows:
\[
  {\Denot{r}} \Def P
\]
where process $P$, and supporting processes $P_1,\ldots,P_n$ 
are defined as follows:
\begin{eqnarray*}
  P & \Def & \Na{in}(q).\IfCCS (\Sat q \And \phi) \Then P_1 \ElseCCS \Co{out}(\False).P\\
  P_i & \Def & \Co{in}_{s_i}(q_i).\Na{out}_{s_i}(r_i).P_{i+1}\\
  P_{n+1} & \Def & \Co{out}(r).P
\end{eqnarray*}
This definition says that when a query $q$ arrives, the offer $\False$
is returned if the offer does not satisfy the trigger condition of the rule.
Otherwise, sequentially, for each supporting process (representing a sub-provider),
a query is made and an offer is received.  Finally, a top-level offer $r$ is made.

We have not yet defined the queries $q_i$ that are made to the
sub-servers, and the final response $r$.
As a first step, we define variable sets $X$, $X_0$, and $X_i$ (for $1 \leq i \leq n$):
\begin{eqnarray*}
  X   & \Def & \Lang{ct} \cup \bigcup_{1 \leq i \leq m} \Lang{ct_i}\\
  X_0 & \Def & X - \Lang{ct}\\
  X_i & \Def & X - \Lang{ct_i}
\end{eqnarray*}
Variable $X$ consists of the parameters of the top-level service and
its immediate sub-services.  Variable $X_0$ consists of the parameters
of the the sub-services only.

Now we can define the final response $r$, and the queries
$q_i$ to the sub-servers in terms of the responses $r_i$
from the sub-servers.  If $X = \{x_1,\ldots,x_n\}$ is a set
of variables, then we write $\exists X. \phi$ as shorthand
for $\exists x_1,\ldots,x_n. \phi$.
\begin{eqnarray*}
   q_i & \Def & QE(\exists X_i. (q \And \psi \And \bigwedge_{1 \leq j < i} r_i))\\
     r & \Def & QE(\exists X_0. (\psi \And \bigwedge_{1 \leq i \leq n} r_i))
\end{eqnarray*}
Intuitively, the first of these definitions says that the query to the $i^{th}$
sub-provider is a quantifier-free formula expressing that 
Recall that $QE(\phi)$ stands for a formula logically
equivalent to $\phi$ but with quantifiers eliminated.
Also, note that formula $q_i$ is a formula over $ct_i$.

\subsection{Semantics of policy composition}

Informally, the meaning of a composite policy is a process
that inputs a query, sends it to each of the sub-policies,
and then takes the disjunction of the responses.  Formally,
the denotation $\Denot{p_1 \oplus p_2}$ is defined as
follows
\[
  \Denot{p_1 \oplus p_2} \Def \Denot{p_1} \oplus \Denot{p_2}
\]
where process operator $\oplus$ is defined as follows:
\begin{eqnarray*}
  P_1 \oplus P_2 & \Def & (P_3 \mid P'_1 \mid P'_2) \setminus \{\Na{in}_1,\Na{out}_1,\Na{in}_2,\Na{out}_2\}\\
  P_3            & \Def & \Na{in}(q).\Co{in}_1(q).\Co{in}_2(q).\Na{out}_1(r_1).\Na{out}_2(r_2).\Co{out}(r_1 \Or r_2).P_3\\ 
  P'_1           & \Def & P_1[\Na{in}/\Na{in}_1,\Na{out}/\Na{out}_1]\\
  P'_2           & \Def & P_2[\Na{in}/\Na{in}_2,\Na{out}/\Na{out}_2]
\end{eqnarray*}

\subsection{Closing a policy process}

The process denoted by a policy communicates with
sub-services through ports of the form $\Na{in}_s$ and
$\Na{out}_s$, where $s$ is a server name.  We need to also
consider how one constructs a server -- i.e. a process with
only ports $\Na{in}$ and $\Na{out}$ -- from such an ``open
server'', along with a collection of servers that will be
used as sub-negotiators.  Suppose $p$ is a policy that
contains server names $s_1,\ldots,s_n$, and $P_1,\ldots,P_n$
is a collection of server processes.  Then we define
\[
  p(P_1,\ldots,P_n) \Def 
     (\Denot{p} \mid P_1[f_1] \mid \ldots \mid P_n[f_n])\rest{L}
\]
where relabelling function $f_i$ maps $\Na{in}$ to
$\Na{in}_{s_i}$ and $\Na{out}$ to $\Na{out}_{s_i}$, and
label set $L$ is
$\{\Na{in}_{s_1},\Na{out}_{s_1},\ldots,\Na{in}_{s_n},\Na{out}_{s_n}\}$.
The (visible) ports of this process are only $\Na{in}$ and
$\Na{out}$.

\section{Correctness}
\label{sec:properties}

We now look at whether policies behave as expected.  We
focus on one correctness property: First, if sub-negotiators
positively respond to queries, will the negotiator defined
by the policy also do so?  This is a kind of preservation
property, and can also be regarded as an assume/guarantee
condition. By ``positively respond'', we mean that a query
$q$ will be responded to with an offer that logically
``intersects'' with $q$ -- in symbols: $\Sat q \And r$.

\begin{definition}
  A negotiator $s$ over a configuration type $ct$ is {\em
    responsive} if, for all queries $q$ in $\LogicOver{ct}$, it
  responds to query $s$ with an offer $r$, such that
  $\Satisfiable{q \And r}$.
\end{definition}

Responsiveness is a strong property not expected of real
negotiators.  A responsive negotiator is nearly miraculous,
in that it will at least partially satisfy every query.  The
point of defining responsiveness is simply to show that the
server defined by a policy will be responsive if the servers
it uses are, too.

\begin{theorem}
Let $p$ be a policy and $s_1,\ldots,s_n$ be servers of the
appropriate type.  Then if $s_1,\ldots,s_n$ are all responsive, 
so is $p(s_1,\ldots,s_n)$.
\end{theorem}

We shall only sketch the proof here.  The core idea is
illuminating and simple to understand, but the full proof
becomes awkward because of notation.  The proof sketch is by
induction on the structure of policies.  We first consider
the case in which a policy is a single rule.  We need the
following simple fact about first-order logic.

\begin{proposition}
Let $\phi$ and $\psi$ be formulas of first-order logic, such
that variable $x$ does not appear free in $\phi$.  Then
$
  \exists x. (\phi \And \psi)  \IffArrow  \phi \And \exists x. \psi.
$
\label{fact:exists-and}
\end{proposition}

\begin{figure}
\begin{center}
\includegraphics[scale=.3]{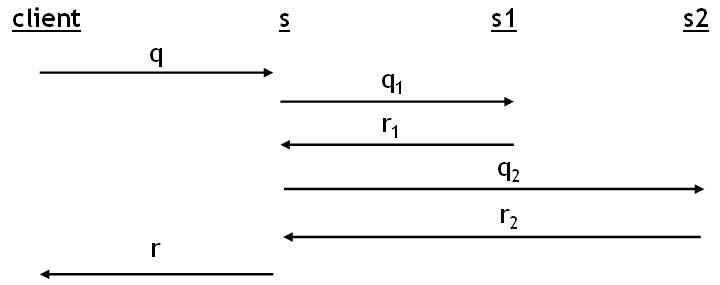}
\end{center}
\caption{Queries and responses of Prop. \ref{correctness}.}
\label{fig:fact-setup}
\end{figure}

Using this fact we can establish a fact about a first-order
logic formula that models a simple policy rule.  Figure
\ref{fig:fact-setup} shows the negotiators involved.  For
simplicity, suppose the config. type of negotiator $s$
concerns only variable $x$, and that the config. types for
negotiators $s_1$ and $s_2$ concern only $x_1$ and $x_2$,
respectively.  Suppose we have a query $q$ (concerning only
$x$), and a formula $\phi$ that has at most variables $x$,
$x_1$, and $x_2$ free.  This formula represents the
conjunction of the policy rule's constraint and the formula
derived from the rule's assignment.  We want to show that if
$\Sat q_1 \And r_1$, and $\Sat q_2 \And r_2$, then $\Sat q
\And r$.

\begin{proposition}
  Suppose $q$ is a formula containing at most $x$ free, $\phi$
  is a formula containing at most $x$, $x_1$, and $x_2$
  free, andformulas $q_2$ and $r$ are defined as follows:
\begin{eqnarray*}
  q_2 & \Def & \exists x,x_1. (q \And \phi \And r_1)\\
    r & \Def & \exists x_1,x_2. (r_1 \And r_2)
\end{eqnarray*}
and 
$
  \Sat \exists x_2. (q_2 \And r_2).
$ Then
$
  \Sat \exists x. (q \And r).
$
\label{correctness}
\end{proposition}

This proposition may seem to be missing the assumption
that $\Sat \exists x_1. q_1 \And r_1$, capturing that
the first negotiation server is responsive.  It is a little
surprising that this assumption is not needed.  The
proof of the proposition is simple.  We have
\[
\begin{array}{llll}
  \Sat & \exists x_2. q_2 \And r_2 & \hspace{2mm} \mbox{(assumption)}\\
  \Sat & \exists x_2. (\exists x,x_1. q \And \phi \And r_1) \And r_2&\hspace{2mm} \mbox{(def. of $q_2$)}\\
  \Sat & \exists x,x_1,x_2. (q \And \phi \And r_1 \And r_2)& 
     \hspace{2mm} \mbox{(Prop.~\ref{fact:exists-and})}\\
  \Sat & \exists x. (q \And \exists x_1,x_2. r_1 \And r_2 \And \phi)&
     \hspace{2mm} \mbox{(Prop.~\ref{fact:exists-and})}\\
  \Sat & \exists x. (q \And r) &\hspace{2mm} \mbox{(def. of $r$)}
\end{array}
\]
which proves the proposition.

We now consider the second case of the proof sketch, in
which a policy is the composition of two policies.
Suppose the query is $q$, and the formula from the response
of the first policy is $r_1$, and the formula from the second
is $r_2$.  By the def. of policy composition, we need to 
show that $\Sat q \And (r_1 \Or r_2)$.  By induction we
assume $\Sat q \And r_1$ and $\Sat q \And r_2$ and from
this it is trivial to show that $\Sat q \And (r_1 \Or r_2)$.
(End of proof sketch.)

\section{Extensions to the Language}
\label{sec:extensions}

\subsection{Preferences}
\label{sec:preferences}

Offers can sometimes simplified by taking into account known
preferences of a clients.  For example, suppose this offer
is computed by a negotiation server:
\[
  \Param{speed} = 6 \And \Param{price} \leq 10 \And \Param{price} \geq 8
\]
Clients seek lowest-priced offers, so the upper-bound is not
useful, and the offer should be simplified to $\Param{speed} =
6 \And \Param{price} = 8$.

To support such simplifications, we need a way to express
client preferences in policy, and logical manipulation to
support these preferences.  For the language, we can
introduce two new clauses at the policy level, just after
the \KWord{serves} clause: \KWord{maximize} and
\KWord{minimize}.  Each takes a list of parameters of the
configuration type served by the policy.  The parameters
listed after \KWord{maximize} are ones the {\em client}
seeks to maximize, and similarly for \KWord{minimize}.
For example in the example policy of Fig.~\ref{fig:storage-example-1},
we could write \verb+minimize: [price]+.

For the logical handling, if $x$ is the parameter that the
client seeks to minimize, and $\phi$ is a condition with $x$
free, we can write the following.
\[
   \phi \And \neg(\exists x'. \phi[x'/x] \And x' < x).
\]
This formula strengthens a condition on $x$ by saying
there exists no value $x'$, which must satisfy the
same condition as $x$ does, but that is less than $x$.
The second conjunct is in $L$, so the existential
quantifier can be eliminated.  

For example, consider again the formula
$
  \Param{speed} = 6 \And \Param{price} \leq 10 \And \Param{price} \geq 8
$.
By the construction we obtain formula
\[
  \Param{s} = 6 \And \Param{p} \leq 10 \And \Param{p} \geq 8
  \And 
  \neg(\exists \Param{p'}. 
  (\Param{s} = 6 \And \Param{p'} \leq 10 \And \Param{p'} \geq 8 \And \Param{p'} < \Param{p}))
\]
The existentially-quantified formula is equivalent to $s = 6 \And p >
8$, so the formula as a whole is equivalent to $s = 6 \And p = 8$.

\subsection{Extended Offers}

In previous sections, an offer from a negotiator has been 
defined as a formula over the configuration type supported 
by the negotiator.  An alternative is to define an offer
as a pair $(r,t)$, where $r$ is a formula as before, and
$t$ is a {\em token}.  The token is what is sometimes called
``opaque'' -- its structure is not visible to a client.

The token serves several purposes.  First, the negotiator
can require that whenever a client accepts an offer, the
token for the offer is supplied.  This is a means to avoid
counterfeit offers.  Second, the token can be used by
the negotiator to record the sub-offers that ``support''
a given offer.  Then, if a client accepts an offer, the
negotiator can use the token to see which previously-received
sub-offers should be accept.
Finally, tokens that capture sub-offers
are helpful in expressing policy correctness properties.
In particular, whether every offer is indeed supported by
sub-offers previously received by the negotiator.

We now briefly outline a form of token that records
sub-offers.  For example, an offer from a home builder might
contain information, not visible to the potential client,
about the offers made by electrical and plumbing contractors
used by the home builder in defining its own offer.  We call
such offers ``extended offers''.  The syntax of extended
offers is defined in Fig.~\ref{fig:offer-syntax}.  The
general form of an extended offer is $(\phi, t)$.  The 
specific form that records sub-offers is a $\BNFnonterm{policy-offer}$.

\begin{figure}
\[
\begin{array}{lcl}
 \BNFnonterm{offer}        & \BNFdef & (\phi, \BNFnonterm{token})\\
 \BNFnonterm{policy-offer} & \BNFdef & (\phi, \BNFnonterm{policy-token})\\
 \BNFnonterm{policy-token} & \BNFdef & \BNFnonterm{rule-offer} \BNFsep 
                                       \BNFnonterm{policy-token} + \BNFnonterm{policy-token}\\
 \BNFnonterm{rule-offer}   & \BNFdef & (\phi, (assn, \BNFnonterm{rule-token}))\\
 \BNFnonterm{rule-token}   & \BNFdef & (\BNFnonterm{ct}, \BNFnonterm{offer}) \BNFsep
                                       \BNFnonterm{rule-token} \times \BNFnonterm{rule-token}
\end{array}
\]
\caption{The structure of extended offers, where $\BNFnonterm{token}$ ranges over 
arbitrary strings, $\BNFnonterm{assn}$ is an assignment, $\BNFnonterm{s}$ is a server id,
and $\BNFnonterm{ct}$ is a configuration type.}
\label{fig:offer-syntax}
\end{figure}

To support extended offers, the policy semantics given in
Section~\ref{sec:interpretation} must be appropriately modified.
For example, in the semantics of policy composition, the 
def. of process operator $\oplus$ must be modified so that
the value sent on port $\Co{out}$ is not $r_1 \Or r_2$,
but $(r_1 \Or r_2, t_1 + t_2)$, where the values received
on ports $\Na{out}_1$ and $\Na{out}_2$ are $t_1$ and $t_2$,
respectively.

Finally, we briefly mention how extended offers can be
used to help capture a correctness condition of policies.
Suppose a negotiator defined by policy provides an
extended policy offer $po$.  Then we expect the following
to hold:
\begin{enumerate}
\item If $po$ has the form $(\phi, (\phi_1,t_1) + \cdots + (\phi_n,t_n))$,
then $\phi \IffArrow \phi_1 \Or \cdots \Or phi_n$.
\item If $po$ contains a rule-offer of the form
$(\phi, (assn, (ct_1, (\phi_1,t_1)) \times \cdots \times (ct_n, (\phi_n,t_n))))$,
and $X_0 \Def \bigcup_{1 \leq i \leq m} \Lang{ct_i}$, 
then $\phi \IffArrow \exists X_0. (\phi_1 \And \cdots \And \phi_n \And \phi_{assn})$.
\end{enumerate}
where $\phi_{assn}$ means the logical condition derived from
assignment $assn$.


\subsection{Parallel Queries to Sub-negotiators}
\label{sec:parallel-queries}

The process derived from a policy rule queries the negotiators
that appear in the rule sequentially.  In general this is
required, because constraints may exist over the sub-offers
from these negotiators.  For example, in our storage example,
the combined capacity of the sub-offers is constrained. 
However, the querying of two negotiators of a rule can be
performed in parallel if the rule places no constraints
between the two negotiators.  For example, a storage policy
could be defined that would simply seek 50 GBytes of
storage from each of two servers, with a maximum price.
These queries could be run in parallel.

Let us be more precise about what it means for a rule to
place no constraints between negotiators.  Suppose we have a
formula of $\Logic$ containing variables $x$ and $y$.  If
$\phi \IffArrow (\exists y.\phi) \And (\exists x.\phi)$,
then $\phi$ can be said to express no constraints between
$x$ and $y$.  Intuitively, this equivalence says that
condition $\phi$ on $x$ and $y$ can be fully captured as a
condition on $x$ alone and a condition on $y$ alone.

Using this idea it is straightforward to work out, using the
semantics of policy rules of
Section~\ref{sec:rule-semantics}, whether the servers of a
rule can be split into two groups such that servers of the
two groups can be processed in parallel.  This analysis
depends on quantifications of the formula $q \And \psi$ that
appears in the semantics.  In some cases the parallelization
can be done at compile time, independently of knowledge of
$q$ (except to know that it is a formula over $ct$).


\section{Implementation}
\label{sec:implementation}


We have developed software building blocks to support
service negotiation.  As part of these we have developed
support for parsing and interpreting policies in our
policy language, including Java code to perform 
quantifier elimination for linear inequalities over
the rationals.


Sometimes quantifier elimination is part of a
theorem-proving system, with the aim to show the validity of
a formula.  Our goal is different: to compute a formula that
is logically equivalent to another formula, but without
quantifiers.  The practical impact of this requirement is
that certain methods used with PA over the naturals, which
do not preserve logical equivalence, are not applicable in
our work.  The situation is similar to the use of
Skolemization in theorem proving, which again does not
generally preserve logical equivalence.


Our current implementation uses the Fourier-Motzkin
algorithm \cite{schrijver1998theory} for quantifier
elimination.  To eliminate a quantifier in a formula $\phi$,
we first put $\phi$ into disjunctive normal form, convert
$\phi$ to a system of linear inequalities, eliminate the
variable of interest, and then convert it back to a logical
formula.

Simplification steps are essential in keeping the
generated queries and replies simple.  We perform
simplification both on linear inequalities and on
formulas.  For the simplification of formulas we
developed a simple rewriting system that attempts
to apply rewrite objects throughout the abstract
syntax tree representation of a formula.


An obvious question in the application of logic to service
negotiation is whether it is practical, especially because
quantifier elimination in PA is doubly
exponential in the size of the formula \cite{FR74}.  We have
not yet run experiments, but there are reasons to be
optimistic.  First, we expect that many services will not
have more than one or two dozen parameters in their
configuration types, meaning that the formulas should not be
large.  (We say this in relation to problems like SAT
solving, which is applied to propositional formulas
containing hundreds of thousands, or even millions of
symbols.)  Second, negotiation of a service happens {\em
  before} its use, and we imagine that for most services
negotiation will happen much less frequently than service
use -- although we do also anticipate ``one-shot'' services
(e.g., a high-def, secure video conference call) which are
used only once after being negotiated on.

\section{Related Work}
\label{sec:related}

There is much work in service negotiation, and the concept
of negotiation performed through a hierarchy of agents is
not new.  For example, see the work on SNAP in \cite{CF02}.
There is also existing work on negotiation policy languages;
for example, see \cite{LBK06} and \cite{GLD03}.  However,
these languages are not based on a hierarchical negotiation
model.  We know of no other work on negotiation policy 
designed to support hierarchical negotiation.

There is also much work on the automation of service
composition and service selection.  For example, see
\cite{BDSN02,IRG06}.  It is important
to understand the difference between that work and
the work presented here.  In this work {\em we do not
compose services} -- we compose negotiations.  
One can understand a rule in our policy language 
as reflecting that a specific implementation of a
service is the composition of other (sub-) services.
The point of the policy negotiation is to make
sure that compatible variants of these sub-services
are obtained, and that the negotiated offer to the
client reflects the offers from the needed sub-services.
On the other hand, a rule of our policy language does
not express {\em how} the sub-services that are being
negotiated for can be put together to form a service.

\section{Conclusions}
\label{sec:conclusion}

We have seen how, by expressing client query and server
offers in logic, and by using quantifier elimination, it is
possible to support hierarchical negotiation using a simple
negotiation policy language.  A service provider, to define
its negotiation strategy, must specify in policy only the
sub-negotiators to be used, and how negotiable parameters of
the service being negotiated on can be defined in terms of
parameters the negotiated service will need to use.  Work
remains to be done to understand the range of services for
which this approach to negotiation is practical.

\subsubsection*{Acknowledgments}

We thank Michael Benedikt for pointing us to quantifier
elimination as a logical means for projecting relations,
and Alan Jeffrey for suggesting that sets of offers be
expressed as logical conditions, and also for other
helpful discussion on the topic of service negotiation.

\bibliographystyle{plain}


\appendix

\section{Brief Overview of CCS}
\label{app:ccs}

CCS is a process algebra created by Robin Milner \cite{Milner89}.
In CCS, a process is an algebraic term built up from a
collection of operators.  Roughly, one can think of a CCS
term as a textual description of a state machine.

The operator $0$ is the {\em nil} or {\em deadlocked}
process.  This is like a state machine with a single state
and no outgoing transitions.

The operator $.$ is the {\em prefix} operator, which takes
an {\em action} on the left and a process term on the right.
An action is just a name, like $\Na{a}$, or a complimented
name, or {\em co-name}, like $\Co{b}$.  An example of the
prefix operator is $\Na{a}.0$.  This process can perform an
$a$ action and then deadlocks.  We say that $\Na{a}.0$ can
perform an $a$ action and then {\em evolve} to process $0$.
The process $\Na{a}.\Na{b}.0$ can perform an $\Na{a}$ action
and then evolve to process $\Na{b}.0$.  In general, a
process $\alpha.P$, where $\alpha$ is some action, and $P$
is a process, can perform $\alpha$ and evolve to process
$P$.

The operator $+$ is the {\em choice} operator, which takes
two processes.  An example is $\Na{a}.0 + \Na{b}.0$.  This
process can perform either an $\Na{a}$ action or a $\Na{b}$,
and then in either case deadlocks.  Generally, if $P$ can
perform an action $\alpha$ and evolve to process $P'$, then
$P + Q$ can perform $\alpha$ and evolve to process $P'$, and
symmetrically for $Q$.

To provide for ``looping'', recursive process definitions
are allowed.  For example, one can define $A \Def \Na{a}.P$.
Process $A$ can repeatedly perform $\Na{a}$ actions.
Generally, if $A \Def P$, and if $P$ can perform action
$\alpha$ and evolve to process $P'$, then $A$ can also
perform action $\alpha$ and evolve to $P'$.  Another example
is $P \Def (\Co{a}.P + \Co{b}.0)$.  Process definitions need
not be recursive.

The names in a process can be changed by using a {\em
  relabelling} function.  The relabelling operator of $CCS$
is written $[f]$, where $f$ is a relabelling function.  For
example, $(\Na{a}.0)[\Na{a}/\Na{b}]$ is exactly like the process
$\Na{b}.0$.  Generally, if $P$ can perform an action
$\alpha$ and evolve to $P'$, then $P[f]$ can perform action
$f(\alpha)$ and evolve to $P'[f]$.

The formal meaning of a CCS process term is given as a {\em
  transition system}, which is a directed graph in which the
nodes are CCS terms and the edges are labelled with actions.
For example, we understand the process $\Na{a}.0$ as a
transition system with nodes $\Na{a}.0$ and $0$, and a
transition from the first to the second, labelled with
$\Na{a}$.  A CCS transition system differs from a finite
state machine: it can have infinitely many nodes, and no
states are marked as end states.

Two CCS process can be put ``in parallel'' using the
{\em parallel composition} operator $\mid$.  For example,
$(\Na{a}.0 + \Co{b}.0) \mid (\Na{a}.0 + \Na{b}.0)$.
When two processes are put in parallel, the resulting
process can behave in two ways.  First, one of the
two components can act {\em independently}.  In the example,
the first component can perform $\Na{a}$, and the
composite evolves to $(0 \mid (\Na{a}.0 + \Na{b}.0)$.
Second, the two components can {\em synchronize}, provided
they can perform complimentary actions.  In the
example, the two components can synchronize on $\Na{b}$
and $\Co{b}$, resulting in distinguished action $\tau$, and
the composite evolves to $(0 \mid 0)$, which incidentally
behaves identically to $0$.

So, in general, if $P$ can perform $\alpha$ and evolve to
$P'$, then $P \mid Q$ can perform $\alpha$ and evolve to $P'
\mid Q$ (and symmetrically for $P$).  Also, if $P$ can
perform $\alpha$ and evolve to $P'$, and $Q$ can perform
$\Co{\alpha}$ and evolve to $Q'$, then $P \mid Q$ can
perform $\tau$ and evolve to $P' \mid Q'$.  Action $\tau$ is
special: it is neither a name nor a co-name, and cannot be
complimented.  It is therefore impossible for $\tau$ actions
to synchronize with other actions.  Intuitively, a $\tau$
action represents activity internal to a system that cannot
be observed outside the system.

The remaining CCS operator is the {\em restriction} operator.
The restriction operator is written $\rest{L}$, where $L$
is a set of non-$\tau$ actions.  Restriction prevents a 
process from performing an action in $L$.  An example is 
$(\Na{a}.0 \mid \Co{a}.0)\rest{\{\Na{a}\}}$. The two 
components of this process can synchronize, but cannot
act independently.  Generally, if $P$ can
perform action $\alpha$ and evolve to process $P'$, and
$\alpha$ and $\Co{\alpha}$ are not in set $L$, then
$P\rest{L}$ can perform $\alpha$ and evolve to $P'\rest{L}$.

\end{document}